\documentclass[aps,prl,superscriptaddress,amsmath,amssymb,floatfix,twocolumn]{revtex4-1}
\usepackage{graphicx}
\usepackage{subfigure}
\usepackage{color}
\usepackage{epstopdf}

\begin{document}

\title{Orbital-selective correlations and renormalized electronic structure in LiFeAs}

\author{Huihang Lin}
\affiliation{Department of Physics and Beijing Key Laboratory of Opto-electronic Functional Materials and Micro-nano Devices,
Renmin University of China, Beijing 100872, China}

\author{Rong Yu}
\email{rong.yu@ruc.edu.cn}
\affiliation{Department of Physics and Beijing Key Laboratory of Opto-electronic Functional Materials and Micro-nano Devices,
Renmin University of China, Beijing 100872, China}

\author{Jian-Xin Zhu}
\email{jxzhu@lanl.gov}
\affiliation{Theoretical Division and Center for Integrated Nanotechnologies, Los Alamos National Laboratory, Los Alamos, New Mexico 87545, USA}

\author{Qimiao Si}
\email{qmsi@rice.edu}
\affiliation{Department of Physics \& Astronomy,
Rice Center for Quantum Materials,
Rice University, Houston, Texas 77005,USA}
%\date{\today}

\begin{abstract}
Multiorbital models are important to both the
correlation physics and topological behavior of quantum materials.
LiFeAs is a prototype iron pnictide suitable for indepth investigation of  this issue.
 Its electronic structure is strikingly different from the prediction of the
noninteracting description.
Here, a multiorbital Hubbard model for this compound is studied using a
 $U(1)$ slave spin theory.
We demonstrate a new mechanism for a large change in the size of the Fermi surface, namely, orbital selectivity of the energy-level renormalization cooperating with
its counterpart in
the quasiparticle spectral weight.
Using this effect, we show how the  dominating  features of the electronic structure in LiFeAs are  understood in terms of  the local   correlations alone.
Our results reveal a remarkable degree of universality out of the
 seemingly complex multiorbital building blocks across a broad range of strongly correlated superconductors.
\end{abstract}

\maketitle

%\onecolumngrid

{\it Introduction.~}
Establishing features that are universal
across the different families of strongly correlated systems
and identifying properties that are particular to each family are important
routes towards elucidating these quantum materials. For iron-based superconductors (FeSCs) \cite{Kamihara_JACS_2008},
an important feature that
is of extensive current interest
 is
their multiorbital
behavior
\cite{Johnston_AP_2010, Wang_Sci_2011, Dagotto_RMP_2013, Dai_RMP_2015, Si_NRM_2016, Hirschfeld_CRP_2016, Bascones_CRP_2016, Yi_npjQM_2017}.
It has been recognized that
electron correlations are strongly orbital dependent in many
FeSCs \cite{Yu_PRB_2011,Yin_NP_2011, Yu_PRL_2013, deMedici_PRL_2014, Backes_PRB_2015, Moreo_CP_2019, Yi_PRL_2013, Wang_NC_2014, Ding_PRB_2014, Wang_PRB_2015, Niu_Feng_PRB_2016, Hiraishi_Hosono_2020,Haihu_PRX_2020}.
This strong orbital selectivity not only causes large effective mass enhancements \cite{Liu_PRB_2015} and
a
substantial renormalization
of the electronic structure in the normal
state \cite{Yi_NC_2015} but
also affects the pairing structure of
 the superconducting
state
 \cite{Yu_PRB:2014,Yin_NP:2014, Ong_PNAS:2016,Nica_npjQM_2017}.
Previous studies showed that the Hund's rule coupling between the multiple orbitals plays a crucial role in suppressing
the interorbital
correlations~\cite{Yu_PRB_2011,deMedici_PRL_2014}
 and pushes the system towards a novel orbital-selective Mott phase (OSMP),
 in which the iron $d_{xy}$ orbital is Mott localized while other $3d$ orbitals are still itinerant~\cite{Yu_PRL_2013}.
 In experiments, this OSMP can be accessed from a Fermi-liquid-like metallic phase
 by
 increasing the temperature~\cite{Yi_PRL_2013}.
 More recently,
angular resolved photoemission spectroscopy (ARPES) measurements   \cite{Yi_arXiv_2020}
have implicated an OSMP as a ground state in the iron chalcogenides upon
isovalent doping~\cite{Yi_arXiv_2020},
and observed a
Fermi surface reconstruction associated with the orbital-selective Mott transition (OSMT):
 as  the OSMP
 is approached, the hole
 Fermi
 pocket with a $d_{xy}$ orbital character
 at the Brillouin zone (BZ) center vanishes
 and new features with the
 itinerant $3z^2-r^2$ orbital character emerges near the X point.
 The orbital selective Mott  physics
 is recognized as universal across essentially all the
 iron chalcogenides~\cite{Yi_NC_2015} and actively interplays with
 their
 nematicity and superconductivity
 \cite{Davis_Science_2017, Davis_NM_2018, Bascones_PRB_2017, Yu_PRL_2018, Hu_PRB_2018}.

In other FeSCs, in particular for iron pnictides, the situation
remains open.
On general grounds, one may expect that several factors, including
the effective orbital dependence in the bandwidth and the extent of the
 kinetic interorbital  hybridization
as well as the degree of orbital-degeneracy breaking (crystal level splitting)
may interplay with the Hund's coupling and influence the strength of the orbital selectivity.
It has been proposed that the orbital selectivity is enhanced when
the $3d$
electron occupation number per Fe site
is decreased
from
$n=6$, the case of the parent
compound,
towards $n=5$~\cite{deMedici_PRL_2014,Yu_COSSMS:2013}
 This well accounts for the unusually large effective masses observed in heavily hole doped iron pnictides $A$Fe$_2$As$_2$ ($A$=K, Rb, Cs),
 and prompts a nearby antiferromagnetic ground state in the phase diagram~\cite{Eilers_PRL_2016}
 that differs from the one in
 the
 parent compounds~\cite{Dai_RMP_2015}. For
 the typical iron pnictides with carrier concentrations close to $n=6$,
 there is not yet
 a clear-cut case
 for
orbital selective correlations.

The 111 iron pnictide LiFeAs
presents a promising and pressing case for elucidating the multiorbital correlations
of the iron pnictides with broader importance.
It superconducts in its pristine form~\cite{Wang_SSC_2008,Tapp_PRB_2008},
and there is no static magnetic or nematic order in its phase diagram.
A recent ARPES study~\cite{Miao_PRB_2016}
reveals features that are reminiscent of the behavior of the iron chalcogenides.
In addition, and particularly notably,
 its correlation-induced renormalization to the
electronic structure near the Fermi energy is especially puzzling.

To put the last feature in a general context, we note that,
in general, electron correlations cause mass enhancements, squeezing the bands toward the Fermi level. Without orbital selectivity,
the Fermi surface would
be
unchanged given that the mass enhancement factors are identical in all bands. Surprisingly,
 in some FeSCs, the observed volume of some Fermi pockets are
shrunken
 compared to that from local density approximation (LDA) calculations;
 this reflects the opposite (``blue/red") correlation-induced shifts of the electron
 and hole bands~\cite{Ortenzi_PRL_2009, Lee_PRL_2012}.
 In particular for
 LiFeAs, LDA obtains three hole
 Fermi
 pockets centered at $\Gamma$ point of the BZ. The outermost one has mainly a $d_{xy}$ orbital character,
 and the inner two are dominated by the $d_{xz/yz}$ orbitals.
 By contrast,
 ARPES measurements show that the band
 giving the innermost hole pocket in LDA is actually below the Fermi
  level~\cite{Borisenko_NP_2016,Brouet_PRB_2016,Fink_PRB_2019,Day_PRL_2018}.
The origin of the
the Fermi-pocket reduction
 is
a subject of controversy.
 In calculations using two-particle self-consistent
  and/or random phase approximation
 approaches~\cite{Zantout_PRL_2019,Hirschfeld_PRB_2020},
 this
 reduction
 is attributed to non-local electron correlations.
 A more recent study~\cite{Kim_Kotliar_2020} suggests that experimental spectrum observed by ARPES can be fit  in terms of a local ({\it i.e.}, {\bf k}-independent)
 self-energy.

\begin{figure}[t!]
\centering\includegraphics[%scale=0.28
width=80mm, trim=5 130 5 20,clip %left, down, right, up
]{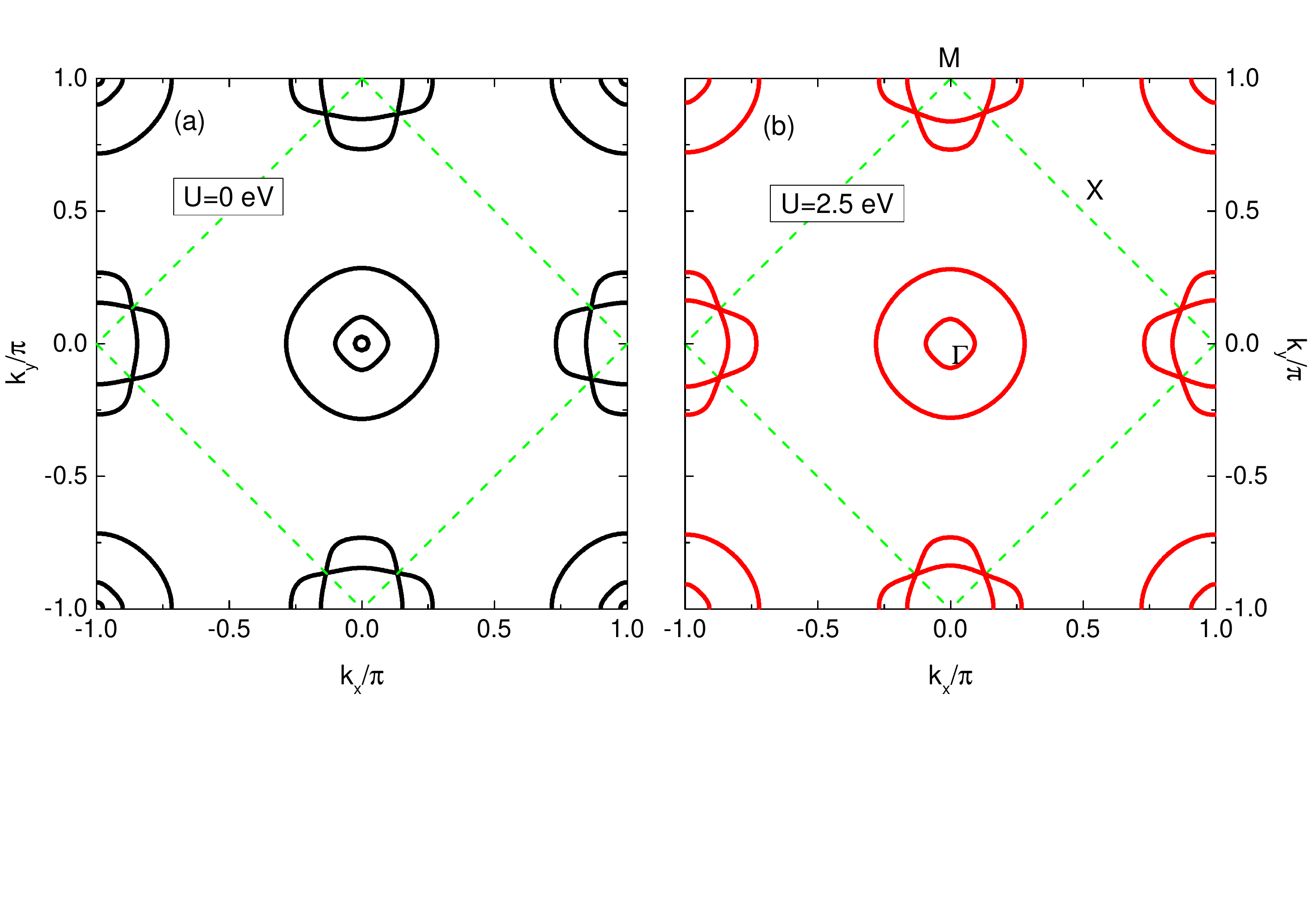}
\caption{(Color online)
Shrinkage of
the
Fermi surface
induced by local correlations in the model for LiFeAs.
The Fermi surfaces are calculated at $U=0$ (in (a)) and $U=2.5$ eV (in (b)), respectively.
In the correlated case (b),
the
 innermost hole pocket
 that appears in the $U=0$ case of
 (a) disappears and all
 the
 other pockets
 are somewhat reduced.
}
\label{fig:1}
\end{figure}

In this letter, we
address
the orbital selectivity of
LiFeAs,
with
two findings. First,
the small As-Fe-As bond angle
 helps
stabilize
an OSMP
over a broad
parameter
regime in the ground-state phase diagram.
With
increasing temperature, the system is driven
 through an OSMT.
Second,
we advance a
natural
but surprising
mechanism for a large change in
 size
of the Fermi surface. We demonstrate a new effect,
an orbital-selective energy-level renormalization, and show
how it cooperates with the orbital dependence in the quasiparticle spectral weight
to cause
a shrinkage of the Fermi
pockets; most drastically,
 the innermost hole pocket
 disappears completely (see Fig.~\ref{fig:1}).
Thus, the
main features of the electronic structure
are captured by
the local correlations.

{\it Model and method.~} We study a five-orbital Hubbard model for LiFeAs. The Hamiltonian reads as
\begin{equation}\label{HamTot}
 H = H_{\rm{TB}} + H_{\rm{soc}} + H_{\rm{int}}.
\end{equation}
$H_{\rm{TB}}$ is a five-orbital tight-binding Hamiltonian with tetragonal lattice symmetry~\cite{Graser_NJP_2009},
\begin{equation}
 \label{Eq:Ham_0} H_{\rm{TB}}=\frac{1}{2}\sum_{ij\alpha\beta\sigma} t^{\alpha\beta}_{ij}
 d^\dagger_{i\alpha\sigma} d_{j\beta\sigma} + \sum_{i\alpha\sigma} (\epsilon_\alpha-\mu) d^\dagger_{i\alpha\sigma} d_{i\alpha\sigma},
\end{equation}
where $d^\dagger_{i\alpha\sigma}$ creates an electron in orbital $\alpha$ ($\alpha=1,...,5$ denoting $xz$, $yz,$ $x^2-y^2$, $xy$, and $3z^2-r^2$ orbitals, respectively) with spin $\sigma$ at site $i$, $\epsilon_\alpha$
refers to the energy level associated with the crystal field splitting (which is diagonal in the orbital basis),
and $\mu$ is the chemical potential
that fixes
 the total electron density to $6$ per Fe.
 Importantly, the inter-orbital hopping
 terms act as a kinetic inter-orbital  hybridization~\cite{Yu_PRB_2017}.
The tight-binding parameters $t^{\alpha\beta}_{ij}$ and $\epsilon_\alpha$ for LiFeAs are presented in the Supplemental Material (SM)~\cite{SM},
which
are determined by fitting the LDA band structure. As shown in Fig.~\ref{fig:1}(a)
[and also, see below,
Fig.~\ref{fig:4}(a)], this model captures major features of the
non-interacting
electronic structure of LiFeAs and gives the correct geometry of
the
LDA
Fermi surface.
$H_{\rm{soc}}= \frac{\lambda^0_{\rm{soc}}}{2} \sum_{i\alpha\beta \sigma \sigma^{\prime}} \left(\mathbf{L}\cdot\boldsymbol{\tau}\right)_{\alpha \sigma, \beta \sigma^{\prime}} d^\dagger_{i\alpha \sigma} d_{i\beta \sigma^{\prime}}$ is an atomic SOC term,
where $\mathbf{L}$ denotes the orbital angular momentum operator and $\boldsymbol{\tau}$ refers to the Pauli matrices.
 $\lambda^0_{\rm{soc}}$ is the (bare) value of the SOC strength without taking into account the effect of electron correlations.
We
take  $\lambda^0_{\rm{soc}}=-30$ meV (see below).
The band splitting caused by the SOC is further renormalized
by the interactions.
The on-site interaction $H_{\rm{int}}$ reads
\begin{eqnarray}
 \label{Eq:Ham_int} H_{\rm{int}} &=& \frac{U}{2} \sum_{i,\alpha,\sigma}n_{i\alpha\sigma}n_{i\alpha\bar{\sigma}}\nonumber\\
 &+&\sum_{i,\alpha<\beta,\sigma} \left\{ U^\prime n_{i\alpha\sigma} n_{i\beta\bar{\sigma}}\right.
 + (U^\prime-J_{\rm{H}}) n_{i\alpha\sigma} n_{i\beta\sigma}\nonumber\\
&-&\left.J_{\rm{H}}(d^\dagger_{i\alpha\sigma}d_{i\alpha\bar{\sigma}} d^\dagger_{i\beta\bar{\sigma}}d_{i\beta\sigma}
 +d^\dagger_{i\alpha\sigma}d^\dagger_{i\alpha\bar{\sigma}}
 d_{i\beta\sigma}d_{i\beta\bar{\sigma}}) \right\},
\end{eqnarray}
where $n_{i\alpha\sigma}=d^\dagger_{i\alpha\sigma} d_{i\alpha\sigma}$.
Here,
$U$, $U^\prime$, and $J_{\rm{H}}$, respectively denote the intra-
and inter- orbital repulsion and the Hund's rule coupling, and
$U^\prime=U-2J_{\rm{H}}$
is taken~\cite{Castellani_PRB_1978}.

We investigate
the
correlation effects
 using a $U(1)$ slave-spin theory~\cite{Yu_PRB_2012, Yu_PRB_2017}.
In this approach, we rewrite $d^\dagger_{i\alpha\sigma} = S^+_{i\alpha\sigma} f^\dagger_{i\alpha\sigma}$,
where $S^+_{i\alpha\sigma}$ ($f^\dagger_{i\alpha\sigma}$) is the introduced quantum $S=1/2$ spin (fermionic spinon)
operator to carry the charge (spin) degree of freedom of the electron;
 the local constraint
 $S^z_{i\alpha\sigma} = f^\dagger_{i\alpha\sigma} f_{i\alpha\sigma} - \frac{1}{2}$.
At the saddle-point level,
the constraint is handled via
 a Lagrange multiplier $\lambda_{\alpha}$ and
the slave-spin and spinon operators
are decomposed
so that $\lambda_{\alpha}$ and the quasiparticle spectral weight $Z_\alpha\propto |\langle S^+_{\alpha} \rangle|^2$ are determined self-consistently~\cite{SM}.

\begin{figure}[t!]
\centering\includegraphics[%scale=0.28
width=85mm]{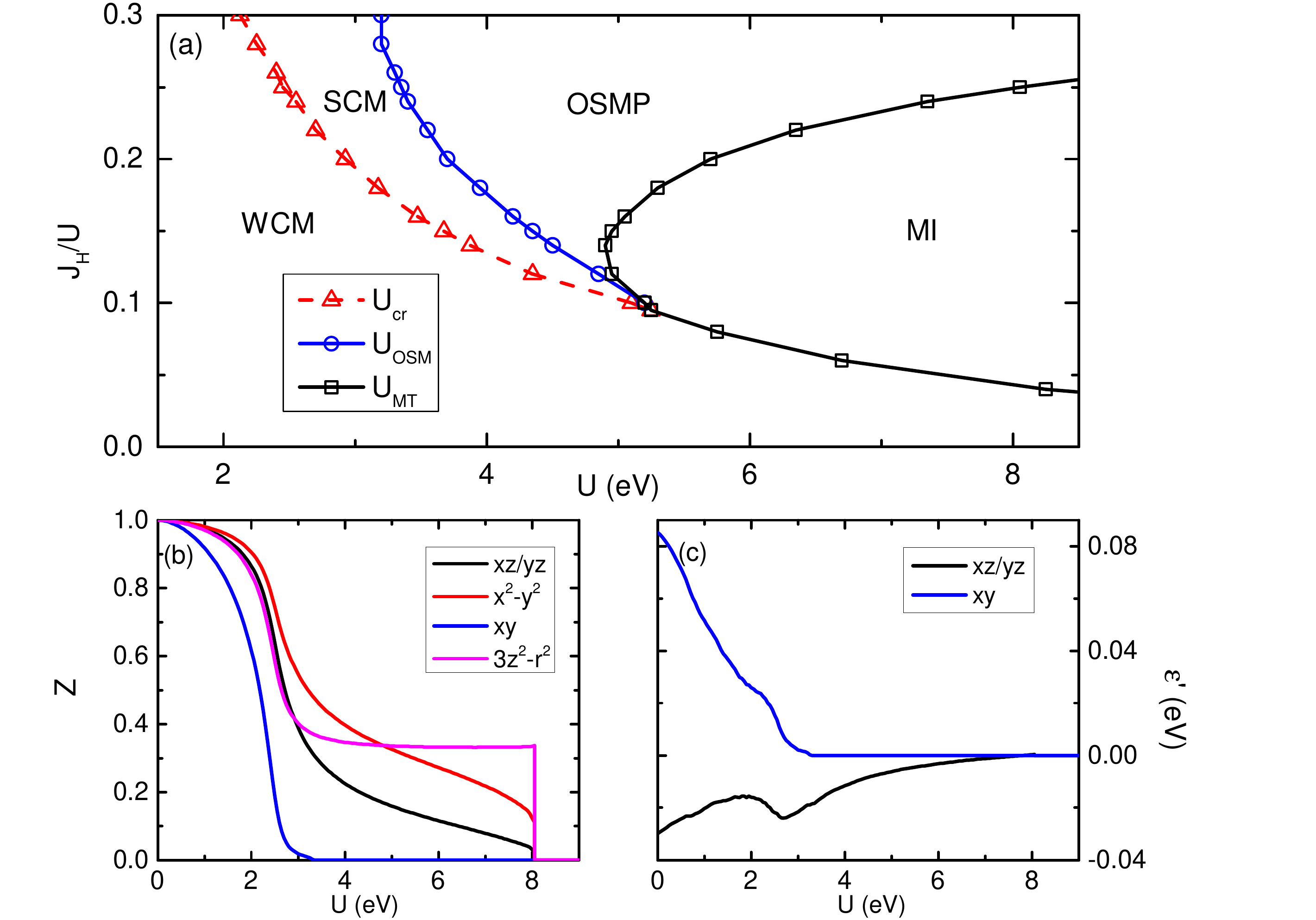}
\caption{(Color online)
(a): Ground-state phase diagrams of the five-orbital Hubbard models for LiFeAs. MI, OSMP, SCM, and WCM denote
the Mott insulator phase,
the orbital-selective Mott phase,
and the regimes of
strongly correlated metal
and weakly correlated metal, respectively.
$U_{\rm{cr}}$ refers to a crossover between WCM and SCM. $U_{\rm{OSM}}$ and $U_{\rm{MT}}$
denote the critical $U$ values of the orbital-selective Mott transition and the Mott transition, respectively.
(b): Evolution of the orbital resolved quasiparticle spectral weights $Z$ with $U$ at $J_{\rm{H}}/U=0.25$ in the model.
(c): Effective level energies of the $d_{xy}$ and $d_{xz/yz}$ orbitals with $U$ at $J_{\rm{H}}/U=0.25$.
}
\label{fig:2}
\end{figure}

{\it Orbital selectivity in the ground state.~} We first
examine
the correlation effects of the model by presenting its ground-state phase diagram in Fig.~\ref{fig:2}(a).
It consists of
a metallic phase, an OSMP, and
a Mott insulator (MI),
with increasing the on-site Coulomb repulsion $U$. In the metallic phase,
 there is a crossover at $U_{\rm{cr}}$ (the red dashed line) from a weakly correlated metal (WCM) to a strongly correlated metal (SCM) with strong orbital selectivity, as shown in Fig.~\ref{fig:2}(b).
In the SCM, the system exhibits bad metal behavior
and $Z_{xy}$ is reduced
  the most.
  $Z_{xy}$ can be suppressed to zero at $U_{\rm{OSM}}$,
  signaling a transition to an OSMP. Further increasing $U$ the system eventually
becomes a MI with electrons in all orbitals localized.

  \begin{figure}[b!]
\centering\includegraphics[%scale=0.28
width=80mm]{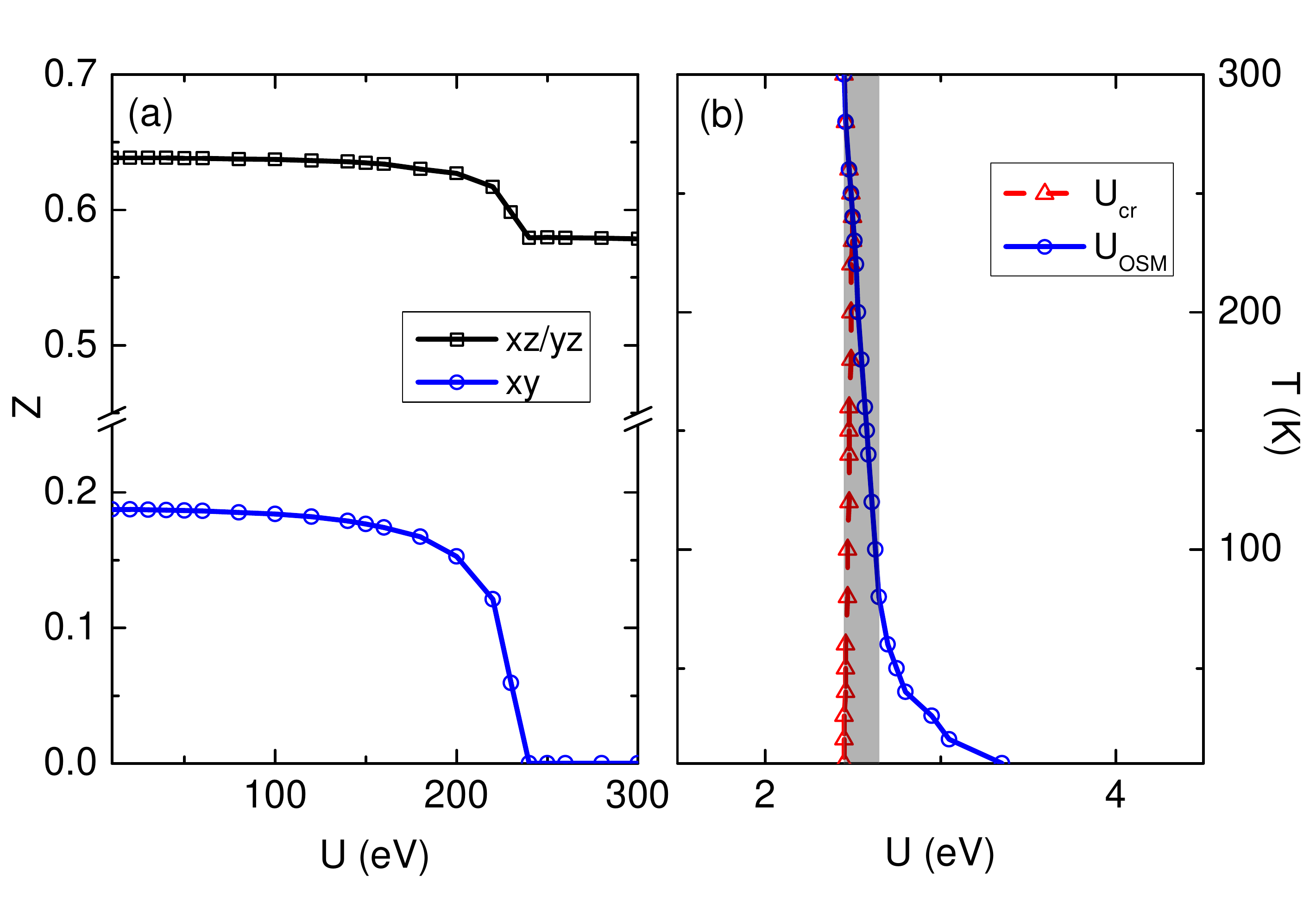}
\caption{(Color online) (a): Evolution of the orbital resolved quasiparticle spectral weights with increasing temperature at
$J_{\rm{H}}/U=0.25$ and $U=2.5$ eV for LiFeAs. An OSMT takes place at about $240$ K. (b): The thermal phase diagram
of the model for LiFeAs at $J_{\rm{H}}/U=0.25$. The shading shows the regime of the estimated $U$ values in LiFeAs.
}
\label{fig:3}
\end{figure}

The phase diagram of LiFeAs
differs qualitatively from those of other parent iron pnictides, such as LaOFeAs, where
no OSMP exists in the ground-state phase diagram
\cite{Yu_PRB_2012}. The stronger orbital selectivity in LiFeAs lies in its smaller As-Fe-As bond angle, which significantly reduces the interorbital hoppings
 involving the $d_{xy}$ orbital (see SM~\cite{SM}).

{\it Temperature induced
orbital-selective
 Mott transition.~}
 Given the proximity to the OSMP in the ground state of LiFeAs,
we address whether this OSMP can be
 approached by increasing the temperature.
In Fig.~\ref{fig:3}(a) we show the evolutions of $Z_{xy}$ and $Z_{xz/yz}$ with temperature
 at $U=2.5$ eV and $J_{\rm{H}}/U=0.25$. Though $Z_{xz/yz}$ drops slightly and keeps to be finite, $Z_{xy}$
 decreases rapidly and vanishes at $T\approx 230$ K, signaling an OSMT. The thermal phase diagram at $J_{\rm{H}}/U=0.25$
 is presented in Fig.~\ref{fig:3}(b). The critical $U$ value for the OSMT, $U_{\rm{OSM}}$, decreases with increasing temperature
 and merges to the crossover line at $U_{\rm{cr}}$ at high temperatures. By comparing to the experimental mass enhancement factors,
 we estimate $U\sim 2.4-2.7$ eV for $J_{\rm{H}}/U=0.25$ (shaded area in Fig.~\ref{fig:4}(b)), which results in an OSMT
 at $T\sim 150-250$ K. This is consistent with the temperature
 evolution
  observed in a recent ARPES experiment~\cite{Miao_PRB_2016}.

{\it Renormalization of electronic structure and shrinkage of Fermi pockets.~}
In a single-orbital model,
electron correlations modify the electronic structure by renormalizing the bandwidth
 while keeping the Fermi surface unchanged as a consequence of
 Luttinger theorem. In a multiorbital model with orbital selectivity,
with
$Z$
being
orbital dependent,
 the situation is
 considerably richer.

Our key finding is
 that the
renormalization of the energy levels, $\epsilon_{\alpha}$,
is also orbital-dependent. This warrants the introduction of a second
renormalization factor, $Z_\alpha^\prime$ that is also
 orbital-dependent, specifying the energy-level renormalization:
\begin{equation}
\label{Eq:Z-prime}
\epsilon_\alpha^\prime=Z_\alpha^\prime \epsilon_\alpha \, .
\end{equation}
The renormalized energy level, $\epsilon_\alpha^\prime$, as a function of $U$ is
 shown in Fig.~\ref{fig:2}(c).
In general, $Z_\alpha^\prime\neq Z_\alpha$, but
is a
 function of all $Z_\alpha$'s;
 it is
 determined
 consistently
from the slave-spin theory.

We now show that the orbital-selective energy-level renormalization,
$Z_\alpha^\prime$, can lead to
a shrinkage
of
the
Fermi
pockets.
We start by specifying the orbital dependence of
the quasiparticle weight $Z$,
which differentiates the renormalization of
the bands with different orbital characters.
This is clearly seen in the calculated bandstructure of LiFeAs in Fig.~\ref{fig:4}: The bottom of the $\gamma$ band,
with a dominant $d_{xy}$ orbital character, is renormalized from about $-0.5$ eV to $-0.1$ eV,
with a renormalization factor of about $5$; while the bottom of the $\delta$ band, with a $d_{xz/yz}$ orbital character,
is only renormalized by a factor of about $1.4$.

We proceed by focusing
on the $\alpha$, $\alpha^\prime$,
and $\beta$ bands near the $\Gamma$ point, which mainly have $d_{xz/yz}$ and $d_{xy}$ orbital characters,
respectively. Here,
 the interorbital hybridization can be neglected.
The dispersion can then be expressed as
\begin{equation}\label{Eq:RenBand}
E(\mathbf{k})\approx Z_\alpha \xi_{\alpha}(\mathbf{k})+\epsilon_\alpha^\prime
= Z_\alpha [\xi_{\alpha}(\mathbf{k})+\epsilon_\alpha] + (Z_\alpha^\prime-Z_\alpha)\epsilon_\alpha,
\end{equation}
where $\xi_\alpha(\mathbf{k})$ is the Fourier component of the hopping parameter $t_{ij}^{\alpha\alpha}$.
For simplicity, we have set the Fermi level to $E=0$, and the Fermi surface in the noninteracting limit is
defined by $\xi_{\alpha}(\mathbf{k})+\epsilon_\alpha=0$. Clearly, the Fermi surface
is un-renormalized
if $Z_\alpha^\prime=Z_\alpha$.
Importantly,
 for
$Z_\alpha^\prime\neq Z_\alpha$, the Fermi pocket either shrinks or
 expands depending on the sign of the last term in Eq.~\eqref{Eq:RenBand}. For LiFeAs, taking $J_{\rm{H}}/U=0.25$
 and $U=2.5$ eV, we find from Fig.~\ref{fig:2} that
$Z^\prime_{xy}\approx0.18<Z_{xy}\approx0.19$, $\epsilon_{xy}>0$,
and $Z^\prime_{xz/yz}\approx0.73>Z_{xz/yz}\approx0.63$,
$\epsilon_{xz/yz}<0$. Therefore, according to Eq.~\eqref{Eq:RenBand},
the pockets of the $\alpha$, $\alpha^\prime$, and $\beta$ sheets
are all reduced in size.
The difference between $Z_{xy}$ and $Z^\prime_{xy}$
is relatively small,
the
Fermi-surface reduction of the $\beta$ sheet is,
correspondingly,
relatively small.
However, the difference between $Z_{xz/yz}$ and $Z^\prime_{xz/yz}$
is larger. Correspondingly, the
shrinkage is
considerably
stronger
 for the inner hole pockets. As shown in Fig.~\ref{fig:1} and Fig.~\ref{fig:4},
 the innermost $\alpha^\prime$ pocket
is
completely eliminated.
Note that the electron pockets also slightly shrink in a way to fulfill the Luttinger theorem.

\begin{figure}[t!]
\centering\includegraphics[%scale=0.28
width=80mm]{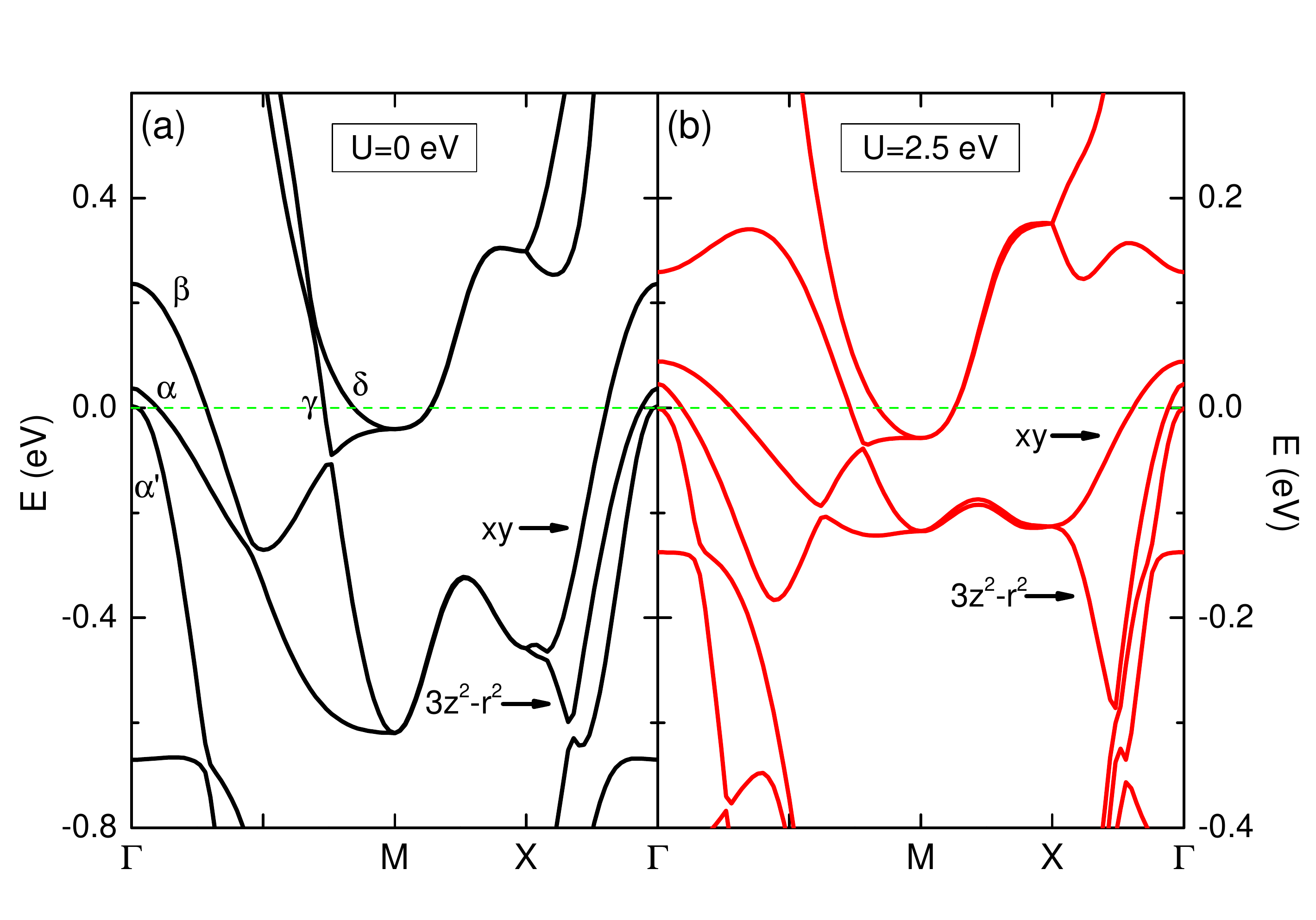}
\caption{(Color online) Comparison of
the
electronic structures in the model for LiFeAs at $U=0$ (in (a))
and $U=2.5$ eV (in (b)), respectively, showing
a
strong renormalization effect.  $J_{\rm{H}}/U=0.25$ and $\lambda^0_{\rm{soc}}=-30$ meV
are taken in the calculation.
}
\label{fig:4}
\end{figure}

{\it Discussions.~}
Several remarks are in order.
 First,
 the calculated low-temperature quasiparticle spectral weights $Z_{xy}\approx0.2$
 and $Z_{xz/yz}\approx0.6$ in the physical regime agree well with the strong orbital dependent mass renormalization factors
 $(m^\star/m_b)_{xy}\sim4-5$ and $(m^\star/m_b)_{xz/yz}\sim1.3-2.3$ found in experiments~\cite{Fink_PRB_2019}.
(We note in passing that
the ARPES line-width
 in LiFeAs may be influenced by the larger $k_z$-dependence
 of the $xz/yz$ orbitals than that of the $xy$ orbital
 \cite{Day_comm}.)
Relatedly,
we show that the system undergoes an OSMT at $T\approx230$ K,
 which explains the strong
 reduction of the
 $xy$-orbital
 spectral weight
  with increasing temperature
  as
   observed in ARPES~\cite{Miao_PRB_2016}.
 Both
 features are to be contrasted with
what happens from the
non-local correlation effects.
The latter would produce a larger renormalization of the quasiparticle weight
to the $3d_{xz/yz}$ orbitals than that to the $3d_{xy}$
orbital~\cite{Zantout_PRL_2019}
given that
the $xz/yz$-orbital-hosting inner hole states have a larger phase space than
the $xy$-orbital-hosting outermost hole Fermi pocket for nested coupling to
the electron Fermi pockets via
 the ($\pi$,$0$)
interactions;
this is opposite to the
ARPES
observations of LiFeAs.
In addition, the
 non-local
mechanism does not account for
 the temperature-induced supression
of the $3d_{xy}$ quasiparticle spectral weight.

Second,
 and importantly, our work advances an entirely new mechanism for
 the Fermi pocket shrinkage
in terms of
local, orbital-selective, electron correlations.
The key new ingredient is our demonstration of the orbital-selective energy-level
renormalization, $Z_\alpha^\prime$
in
Eqs.\,(\ref{Eq:Z-prime},\ref{Eq:RenBand}).
The Fermi surface differs from the one
in the noninteracting limit only when $Z_\alpha^\prime\neq Z_\alpha$, which is necessarily
the case
in the presence of orbital selectivity.
 Our results imply that the
dominant effect of electronic-structure-renormalization in LiFeAs,
\emph{e.g.}
 the
 shrinkage of inner hole pockets,
 is
 understood by the
 local correlations alone.
 More generally, our work advances
 a new mechanism for the general
 phenomenon of
``blue/red shifts"
~\cite{Ortenzi_PRL_2009},
namely the differing renormalization of the energy levels and quasiparticle
 weights,
  $Z_\alpha^\prime\neq Z_\alpha$.

 Third, our results suggest that the electronic structure of the iron pnictides,
 like their iron chalcogenide counterpart,
 is
 predominantly influenced by the local electron correlations.
 Such correlations provide a starting point to understand the nature of the superconducting state \cite{Davis_Science_2017,Davis_NM_2018,Hu_PRB_2018};
indeed, our work sets the stage to address the role of short-range interactions
in driving the superconductivity of LiFeAs and its derivatives, as has been experimentally implicated \cite{Miao_Natcomm_2015,Umezawa_PRL_2012}.

We close the discussion by noting on several additional features.
Because of the strong renormalization of the $xy$ orbital, the
 $3z^2-r^2$ band along the $\Gamma$-X direction is pushed
 closer
 to the Fermi level in LiFeAs, as shown in Fig.~\ref{fig:4}.
 This
 serves a particularly convenient diagnostic of
 strong orbital selectivity~\cite{Yi_arXiv_2020}.
In addition, in
strongly correlated electronic topology,
multiorbital correlations can play a crucial role for
 the band inversion and other inherent multi-band behavior.
LiFeAs is emerging as a candidate system in which  its band topology in the bulk
leads to topologically-nontrivial
superconductivity on its surface~\cite{Zhang_NP_2019}.
Given the important role that strong correlations are expected to play in the band
inversion of Fe-based systems~\cite{Lohani_PRB_2020}, the orbital-selective correlations
we have advanced for LiFeAs set the stage for the much-needed understanding
of the topological behavior in LiFeAs.
In this connection, we note that,
taking into account the renormalization effect on the SOC, the calculated splitting
 between the $\alpha$ and $\alpha^\prime$ bands
 is about $20$ meV (Fig.~\ref{fig:4}(b)), which is consistent with the reported value in
 experiments~\cite{Borisenko_NP_2016, Day_PRL_2018}.

{\it Conclusions.~}
We
have studied
electron correlation effects in a multiorbital Hubbard model for LiFeAs.
An
orbital-selective Mott phase is found to be stabilized in
a broad region of
 the ground-state phase diagram.
 We have identified a new effect of multiorbital correlations,
 namely orbital selectivity in the energy-level renormalization.
 This effect interplays with
 its counterpart in the
 quasiparticle spectral weight, leading to a natural understanding of
 the shrinkage in the
 Fermi pockets and other renormalization of the electronic structure.
Our results suggest that the electronic structure of the iron pnictides, like their iron chalcogenide counterpart,
 is strongly influenced by the  local electron correlations.
 Such correlations underlie the bad-metal normal state
 observed in many families of strongly correlate superconductors.
  Our results not only provide the understanding of
 a striking puzzle in the iron pnictides but also
 uncover a hidden simplicity in the seeming complexity
 of the multiorbital superconductors. This implicates a
 remarkable degree of universality
 across the iron-based superconductors that is shared
 with many other families of strongly correlated superconductors.

%\acknowledgements
\begin{acknowledgments}
We thank M. Yi.
J. W. Huang, A. Damascelli, R. Day, and P. C. Dai
 for useful discussions. This work has in part been supported by
the National Science Foundation of China Grant No. 11674392, Ministry of Science and Technology of China,
National Program on Key Research Project Grant No.2016YFA0300504 and Research Funds of Remnin University
of China Grant No. 18XNLG24 (R.Y. and H.L.),
and by
the U.S. Department of Energy, Office of Science, Basic Energy Sciences, under Award No. DE-SC0018197,
the Robert A.\ Welch Foundation Grant No.\ C-1411 (Q.S.).
Work at Los Alamos was carried was carried out under the auspices of the U.S. DOE NNSA under Contract No. 89233218CNA000001.
It was supported by LANL LDRD Program and in part by the Center for Integrated Nanotechnologies, a U.S. DOE BES user facility.
Q.S. acknowledges the hospitality of the Aspen Center for Physics,
which is supported by NSF grant No. PHY-1607611.
\end{acknowledgments}

%%%%%%%%%%%%%%%%%%%%%%%%%%%%%%%%%%%%%%%%%%%%%

%%%%%%%%%%%%%%%%%%%%%%%%%%%%%%%%%%%%%%%%%%%%%

%%%%%%%%%%%%%%%%%%%%%%%%%%%%%%%%%%
%%%  Supplementary Materials   %%%
%%%%%%%%%%%%%%%%%%%%%%%%%%%%%%%%%%

\newpage
\setcounter{figure}{0}
\makeatletter
\renewcommand{\thefigure}{S\@arabic\c@figure}
\onecolumngrid
\section{ SUPPLEMENTAL MATERIAL -- Orbital-selective correlations and renormalized electronic structure in LiFeAs}
%\newpage

%\onecolumngrid

%\section*{{\Large Supplementary Material for EPAPS}}

\subsection{Details on the tight-binding model}

%\appendix
%\section{Tight-binding parameterization}\label{Sec:AppA}

To obtain the tight-binding
parameters, we perform LDA calculations for LiFeAs, and fit the LDA bandstructure to the
tight-binding Hamiltonian.
We use the form of the
five-orbital tight-binding Hamiltonian given in Ref.~\cite{Graser_NJP_2009}.
The tight-binding parameters so derived are listed in Table~S1.%~\ref{tab:1}.

The electronic structure of the tight-binding model is shown in Fig.~\ref{fig:4}(a), which produces the Fermi surface of three hole pockets centered about $\Gamma$ point and two electron pockets centered about M point, as presented in Fig.~\ref{fig:1}(a).

To clarify the origin of the enhanced orbital selectivity compared to other parent iron pnictides, we calculated the total density of states (DoS) and the DoS projected to the $d_{xy}$ of the tight-binding model. They are contrasted to those of the tight-binding model for LaOFeAs~\cite{Graser_NJP_2009} in Fig.~\ref{fig:S1}. Though the overall DoS of the two models are comparable, the DoS projected to the $d_{xy}$ orbital in the model of LiFeAs has a much narrower bandwidth ($\sim 2$ eV) than in LaOFeAs ($\sim 3$ eV). This indeed suggests a stronger orbital selectivity in
LiFeAs; see the
main text for
a
detailed discussion.

\subsection{Details on the $U(1)$ slave-spin theory}

Here we present a brief introduction to the $U(1)$ slave-spin method. For more details, we refer to Refs.~\cite{Yu_PRB_2012,Yu_PRB_2017}.

In the $U(1)$ slave-spin formulation, we introduce a quantum $S=1/2$ spin operator
whose XY component ($S^+_{i\alpha\sigma}$) is used to represent
the charge degree of freedom of the electron at each site $i$, in each orbital $\alpha$
and for each spin flavor $\sigma$. Correspondingly, we
introduce a fermionic ``spinon'' operator
($f^\dagger_{i\alpha\sigma}$) to carry the spin degree of freedom.
The electron creation operator is
represented as follows,
\begin{equation}\tag{S1}
 \label{Eq:SSCreate}
 d^\dagger_{i\alpha\sigma} = S^+_{i\alpha\sigma} f^\dagger_{i\alpha\sigma}.
\end{equation}
This
representation has an enlarged Hilbert space
compared to
 the
 one for the physical $d$ electrons. To
 restrict
 the Hilbert space to the physical one, we implement a local constraint,
\begin{equation}\tag{S2}
 \label{Eq:constraint}
 S^z_{i\alpha\sigma} = f^\dagger_{i\alpha\sigma} f_{i\alpha\sigma} - \frac{1}{2}.
\end{equation}

This representation contains
a $U(1)$ gauge redundancy corresponding to
$f^\dagger_{i\alpha\sigma}\rightarrow f^\dagger_{i\alpha\sigma} e^{-i\theta_{i\alpha\sigma}}$
and $S^+_{i\alpha\sigma}\rightarrow S^+_{i\alpha\sigma} e^{i\theta_{i\alpha\sigma}}$.
As a result,
 the slave spins can be used to
 carry the
 $U(1)$-symmetric
 physical charge degree of freedom,
 similarly as in the slave-rotor approach~\cite{Florens_Georges_PRB_2004}.

To ensure that the saddle point captures
the correct
quasiparticle spectral weight in the non-interacting
limit (being equal to $1$), we define a dressed operator in the Schwinger
boson representation of the slave spins (in a way similar to the standard
slave-boson theory~\cite{KotliarRuckenstein}):
\begin{equation}\tag{S3}
 \label{Eq:Zdagger}
 \hat{z}^\dagger_{i\alpha\sigma} = P^+_{i\alpha\sigma} S^+_{i\alpha\sigma}
 P^-_{i\alpha\sigma},
\end{equation}
where $P^\pm_{i\alpha\sigma}=1/\sqrt{1/2+\delta \pm S^z_{i\alpha\sigma}}$, and $\delta$ is an infinitesimal positive
number to regulate $P^\pm_{i\alpha\sigma}$.
With this construction,
Eq.~\eqref{Eq:SSCreate} becomes
\begin{equation}\tag{S4}\label{Eq:SBcreate}
d^\dagger_{i\alpha\sigma}=\hat{z}^\dagger_{i\alpha\sigma} f^\dagger_{i\alpha\sigma}.
\end{equation}
The Hamiltonian given in Eq. (1) of the main text
can then be effectively
 rewritten as
%\begin{eqnarray}
\begin{align}\tag{S5}
\label{Eq:HamSS}
H &= \frac{1}{2}\sum_{ij\alpha\beta\sigma} t^{\alpha\beta}_{ij}
 \hat{z}^\dagger_{i\alpha\sigma} \hat{z}_{j\beta\sigma} f^\dagger_{i\alpha\sigma} f_{j\beta\sigma}
 + \sum_{i\alpha\sigma}  (\epsilon_\alpha -\mu) f^\dagger_{i\alpha\sigma}
 f_{i\alpha\sigma}
 \nonumber\\
 & - \lambda_{i\alpha\sigma}[f^\dagger_{i\alpha\sigma}
 f_{i\alpha\sigma}-S^z_{i\alpha\sigma}-1/2] + H^S_{\mathrm{int}}.\nonumber
\end{align}
%\end{eqnarray}
Here, $\lambda_{i\alpha\sigma}$ is a Lagrange multiplier used
 to enforce the constraint in Eq.~\eqref{Eq:constraint}.
In addition,  $H^S_{\mathrm{int}}$ is the interaction Hamiltonian in Eq.~(3) of the main text rewritten in the slave-spin representation
$H_{\mathrm{int}}\rightarrow H_{\mathrm{int}}(\mathbf{S})$~\cite{Yu_PRB_2012}.
The quasiparticle spectral weight
\begin{equation}\tag{S6}
\label{Eq:qpWeightZ}
Z_{i\alpha\sigma}=
|\langle \hat{z}_{i\alpha\sigma}\rangle|^2 \propto
|\langle S^+_{i\alpha\sigma}\rangle|^2 .
\end{equation}

After decomposing the slave spin and spinon operators and treating the constraint on average,
we obtain two saddle-point Hamiltonians for the spinons and the slave spins, respectively:
%\begin{eqnarray}
\begin{align}\tag{S7}
 \label{Eq:Hfmf}
 H^{\mathrm{mf}}_f &=  \sum_{k\alpha\beta}\left[ \xi^{\alpha\beta}_{k} \langle \tilde{z}^\dagger_\alpha \rangle
  \langle \tilde{z}_\beta \rangle + \delta_{\alpha\beta}(\epsilon_\alpha-\lambda_\alpha
  -\mu)\right] f^\dagger_{k\alpha} f_{k\beta},\nonumber\\
 \nonumber\\
 \label{Eq:HSSmf}
 H^{\mathrm{mf}}_{S} &= \sum_{\alpha\beta} \left[Q^f_{\alpha\beta}
 \left(\langle \tilde{z}^\dagger_\alpha\rangle \tilde{z}_\beta+ \langle \tilde{z}_\beta\rangle \tilde{z}^\dagger_\alpha\right)
 + \delta_{\alpha\beta}
 (\lambda_\alpha-\tilde{\mu}_\alpha) S^z_\alpha
 \right]
 + H^S_{\mathrm{int}},\tag{S8}
\end{align}
%\end{eqnarray}
where $\delta_{\alpha\beta}$ is Kronecker's delta function,
$\xi^{\alpha\beta}_{k}=\frac{1}{N}\sum_{ij\sigma} t^{\alpha\beta}_{ij} e^{ik(r_i-r_j)}$, and
%\begin{eqnarray}
\begin{align}\tag{S9}
\label{Eq:Qf}
Q^f_{\alpha\beta} &= \sum_{k\sigma}\xi^{\alpha\beta}_k\langle f^\dagger_{k\alpha\sigma}
f_{k\beta\sigma}\rangle/2,
\end{align}
%\end{eqnarray}
In addition,
$\tilde{\mu}_\alpha$ is an effective onsite potential whose definition is given in Ref.~\cite{Yu_PRB_2012}.

\begin{figure}[h!]
\centering\includegraphics[%scale=0.28
width=100mm%, trim=10 130 25 50,clip %left, down, right, up
]{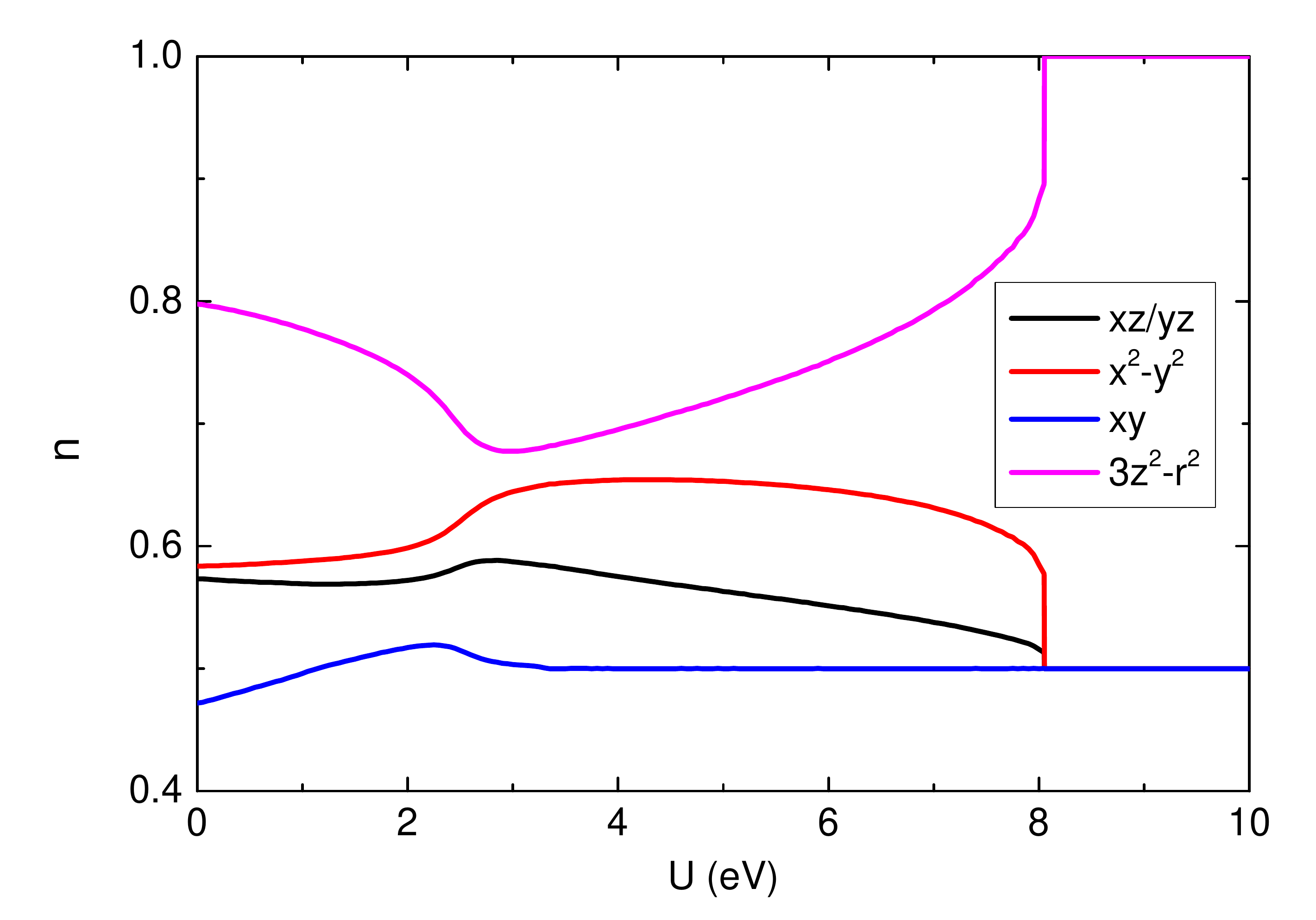}
\caption{(Color online) Evolution of the orbital resolved electron occupation number $n$ with $U$ at $J_{\rm{H}}/U=0.25$ for LiFeAs.
}
\label{fig:S2}
\end{figure}

Eqs.~\eqref{Eq:Hfmf} and \eqref{Eq:HSSmf} represent the main formulation of the $U(1)$
slave-spin approach at the saddle-point level. We study the metal-to-insulator transitions in the paramagnetic phase preserving the translational symmetry. The latter allows us to drop the spin and/or site indices of the slave spins and the Lagrange multiplier $\lambda_\alpha$ in Eqs.~\eqref{Eq:Hfmf} and \eqref{Eq:HSSmf}. We refer to
Refs.\,\cite{Yu_PRB_2012}
and \cite{Yu_PRB_2017}
for a detailed derivation of these
saddle-point Hamiltonians.
At the saddle-point level, $Z_\alpha$ and $\lambda_\alpha$ are solved self-consistently. For a general multiorbital model three saddle-point solutions can be stabilized:
a metallic state with the quasiparticle spectral weight $Z_{\alpha}>0$ in all
the orbitals, a Mott insulator (MI) with $Z_{\alpha}=0$ in all the orbitals with a gapless spinon spectrum, and an OSMP in which
$Z_{\alpha}=0$ in some orbitals whereas $Z_{\alpha}>0$ in the others. The solution of the model for LiFeAs is presented in Fig.~\ref{fig:2},
along with Fig.~\ref{fig:S2}.

Note that in this
approach,
the spinon dispersion
(along with the dispersion of the physical electrons in the metallic phase) is naturally
 renormalized by the quasiparticle spectral weights $\sqrt{Z_\alpha Z_{\beta}}$ and shifted by the effective
 level energy $\epsilon^\prime_\alpha=\epsilon_\alpha-\lambda_\alpha=Z^\prime_\alpha \epsilon_\alpha$, where $Z^\prime_\alpha$ is a complicated function of $Z_\alpha$ and $\lambda_\alpha$ in all orbitals. (Here we set the Fermi level at zero energy.)

\subsection{Bond angle and enhanced orbital selectivity in LiFeAs}

\begin{figure}[h!]
\centering\includegraphics[%scale=0.28
width=100mm%, trim=10 130 25 50,clip %left, down, right, up
]{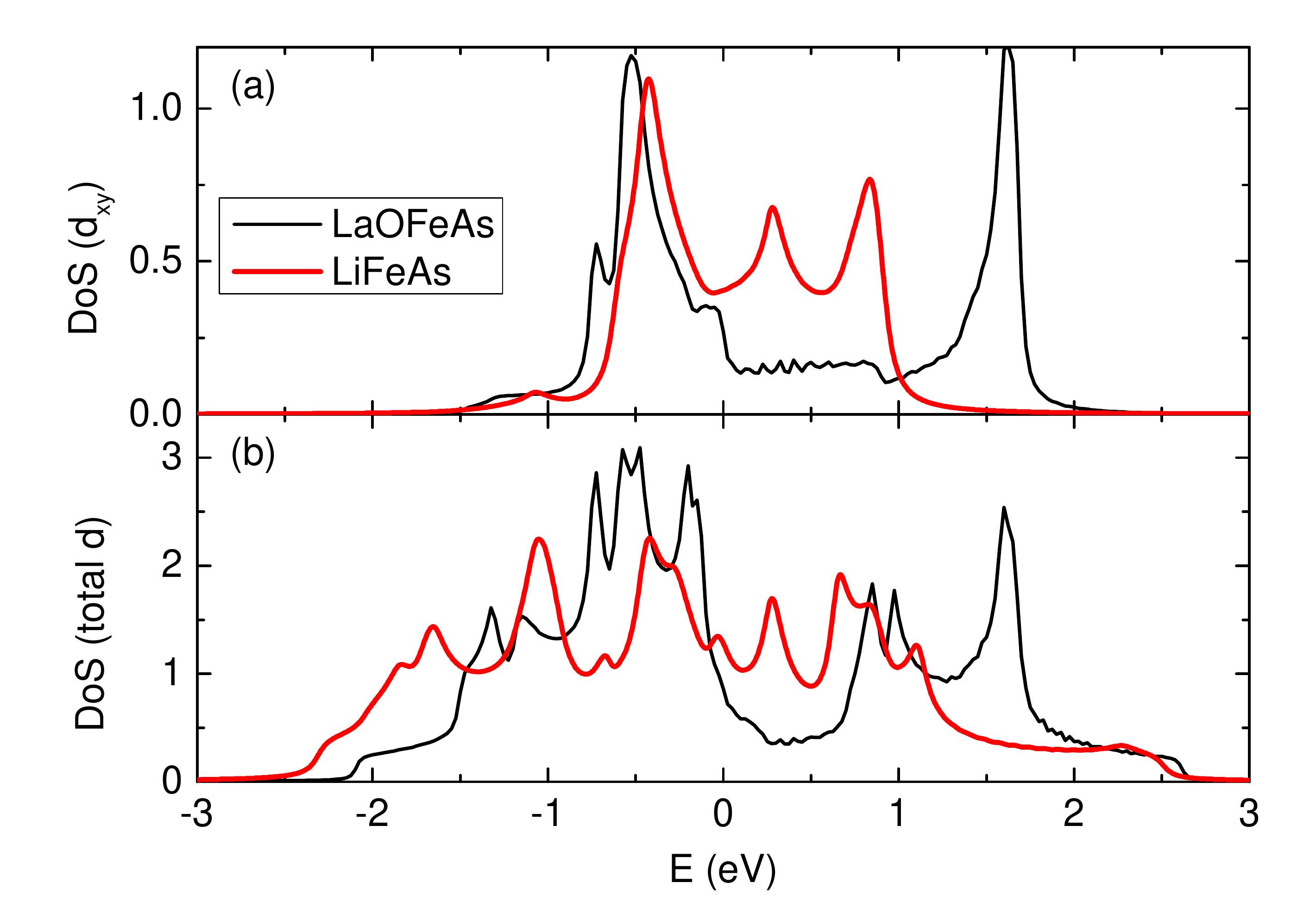}
\caption{(Color online) Density of states (DoS) projected onto the $d_{xy}$ (in (a)) and the total $d$ orbitals (in (b)) at $U=0$ of models for LiFeAs (red curves) and LaOFeAs (black curves), respectively.
}
\label{fig:S1}
\end{figure}

As shown in Fig.~\ref{fig:2}(a) of the main text, the ground-state phase diagram of LiFeAs contains a large regime of OSMP. This differs qualitatively from other parent iron pnictides for which the OSMP was not
stabilized
in the ground state~\cite{Yu_PRB_2012};
it shows that, in the overall phase diagram of the iron pnictides, the OSMP anchors
the SCM phase with a large orbital selectivity.
 To understand this enhanced orbital selectivity in LiFeAs, we note that its As-Fe-As bond angle is about $103^\circ$, which is much smaller than those of other parent iron pnictides ($\gtrsim 110^\circ$), but is similar to that in Li$_{0.8}$Fe$_{0.2}$OHFeSe~\cite{Pitcher_CC_2008, Yi_npjQM_2017}. A smaller bond angle corresponds
 to an elongated tetragon that significantly reduces the interorbital hoppings involving the $d_{xy}$ orbital because
 of the in-plane symmetry of this orbital. This is clearly seen by comparing the interorbital hoppings
 of LiFeAs in Table~S1 to those of LaOFeAs in Ref.~\cite{Graser_NJP_2009}.
 Smaller interorbital hoppings result in narrower bandwidth projected to the $d_{xy}$ orbital.
 As shown in Fig.~\ref{fig:S1}, the
 projected  band  of  the
   $d_{xy}$ orbital in LiFeAs is much narrower than that
 of LaOFeAs though the overall $d$-orbital bandwidths are comparable
 between the two cases. This
 feature leads to the stronger orbital selectivity
 in LiFeAs.

%%%%%%%%%%%%%%%%%%%%%%%%%%%%%%%%%%%%%%%%%%%%%%%%%%%%%%%%%%%%%%%

%%%%%%%%%%%%%%%%%%%%%%%
%\begin{table*}[h!]
%  \centering
\begin{minipage}{\linewidth}
\begin{center}
\begin{tabular}{cccccccc}
  \hline
  % after \\: \hline or \cline{col1-col2} \cline{col3-col4} ...
  \hline
    & $\alpha=1$ & $\alpha=2$ & $\alpha=3$ & $\alpha=4$ & $\alpha=5$ &   &   \\ \hline
  $\epsilon_\alpha$ & 0.06517 & 0.06517 & -0.38097 & 0.18026 & -0.63628 &  & \\ \hline\hline
  $t^{\alpha\alpha}_\mu$ & $\mu=x$ & $\mu=y$ & $\mu=xy$ & $\mu=xx$ & $\mu=xxy$ & $\mu=xyy$ & $\mu=xxyy$ \\ \hline
  $\alpha=1$ & -0.02507 & -0.49888 & 0.24903 & 0.04834 & 0.00813 & -0.02776 & 0.04531 \\ \hline
  $\alpha=3$ & 0.42894 &  & -0.01946 & -0.01032 &  &  &  \\ \hline
  $\alpha=4$ & 0.16275 &  & 0.13559 & -0.00441 & -0.05245 &  & -0.03593 \\ \hline
  $\alpha=5$ & -0.08510 &  &  & -0.04632 & 0.01048 &  & -0.00195 \\ \hline\hline
  $t^{\alpha\beta}_\mu$ & $\mu=x$ & $\mu=xy$ & $\mu=xxy$ & $\mu=xxyy$ &  &  &  \\ \hline
  $\alpha\beta=12$ &  & 0.19318 & -0.05864 & 0.07046 &  &  &  \\ \hline
  $\alpha\beta=13$ & -0.42376 & 0.07714 & 0.01353 &  &  &  &  \\ \hline
  $\alpha\beta=14$ & 0.03406 & -0.02355 & -0.00376 &  &  &  &  \\ \hline
  $\alpha\beta=15$ & -0.14608 & -0.09700 &  & -0.00683 &  &  &  \\ \hline
  $\alpha\beta=34$ &  &  & -0.00635 &  &  &  &  \\ \hline
  $\alpha\beta=35$ & -0.26547 &  & 0.03472 &  &  &  &  \\ \hline
  $\alpha\beta=45$ &  & -0.10611 &  & 0.03363 &  &  &  \\ \hline
  \hline
\end{tabular}
\par
\end{center}
\bigskip
\noindent
{\bf Supplemental Table S}1.
Tight-binding parameters of the five-orbital model for LiFeAs. Here we
use the same notation as in Ref.~\cite{Graser_NJP_2009}. The orbital index $\alpha=$1,2,3,4,5 correspond to $d_{xz}$, $d_{yz}$, $d_{x^2-y^2}$, $d_{xy}$, and $d_{3z^2-r^2}$ orbitals, respectively.
The listed parameters are in eV.
\end{minipage}
%%%%%%%%%%%%%%%%%%%%%%%

\end{document}